\documentclass[a4paper,11pt]{article}
\pdfoutput=1 % if your are submitting a pdflatex (i.e. if you have
\usepackage{jinstpub} % for details on the use of the package, please
\title{\boldmath Spatial profiles of collimated laser Compton-scattering $\gamma$-ray beams}

\author[a]{Takashi Ari-Izumi,}
\author[b,c,1]{Ioana Gheorghe, \note{Corresponding author.}}
\author[b]{Dan Filipescu,}
\author[d]{Satoshi Hashimoto,}
\author[d]{\\Shuji Miyamoto}
\author[a,e]{and Hiroaki Utsunomiya}
\affiliation[a]{Konan University, Department of Physics,\\8-9-1 Okamoto, Higashinada, Japan}
\affiliation[b]{Horia Hulubei National Institute of Physics and Nuclear Engineering -- IFIN-HH,\\30 Reactorului, 077125 Bucharest, Romania}
\affiliation[c]{Department of Physics, University Politehnica of Bucharest,\\Splaiul Independentei 313, 060042 Bucharest, Romania}
\affiliation[d]{Laboratory of Advanced Science and Technology for Industry, University of Hyogo,\\3-1-2 Kouto, Kamigori, Ako-gun, Hyogo 678-1205, Japan}
\affiliation[e]{Shanghai Advanced Research Institute, Chinese Academy of Sciences, \\Shanghai 201210, China}

\emailAdd{ioana.gheorghe@nipne.ro}

\abstract{
The intensity and energy spatial distributions of collimated laser Compton scattering (LCS) $\gamma$-ray beams and of the associated bremsstrahlung beams have been investigated as functions of the electron beam energy, electron beam phase space distribution, laser optics conditions and laser polarization. We show that the beam halo is affected to different extents by variations in the above listed parameters. In the present work, we have used laser Compton scattering simulations performed with the \texttt{eliLaBr} code (https://github.com/dan-mihai-filipescu/eliLaBr) and real LCS and bremsstrahlung $\gamma$-ray beams produced at the NewSUBARU synchrotron radiation facility. A 500~$\mu$m MiniPIX X-ray camera was used as beamspot monitor in a wide $\gamma$-ray beam energy range between 1.73~MeV and 38.1~MeV.  

}


\begin{document}
\maketitle
\flushbottom

\section{Introduction}
\label{label_sec_intro}

High directivity, intense flux and narrow spatial and energy distributions make laser Compton scattering (LCS) $\gamma$-ray beams suitable tools for nuclear applications with costly target materials. Due to the small beam spot size, less target material is necessary for producing the required specific activities, which is critical for medical radioisotopes production~\cite{HabsKoster2011,WantingPan_WenLuo2021}, as well as for low energy and low cross section measurements of ($\gamma$,~$\gamma'$) and ($\gamma$,~particle) reactions for nuclear astrophysics~\cite{Arnold2012,Utsunomiya2015,Filipescu2014,Nyhus2014} and nuclear structure studies~\cite{Isaak2011,Derya2014}. The use of $\gamma$-ray beams with small opening angle is also advantageous for absolute reactions cross sections measurements, as compact targets of a few mm diameter are enough for the entire photon flux to be impinged on the target surface. Here we note the extended campaign of photoneutron~\cite{Gheorghe2017,krzysiek2019_ho165,Kawano2020,Gheorghe_ND2022} and photofission~\cite{Filipescu_ND2022} cross section measurements in the Giant Dipole Resonance energy region conducted during 2015~--~2019 at the LCS $\gamma$-ray beam line of the NewSUBARU synchrotron facility as part of the recent IAEA Coordinate Research Project on Updating the Photonuclear Data Library and generating a Reference Database for Photon Strength Functions~\cite{CRP_website,Goriely2018}. 

It is well known that the LCS $\gamma$-ray beam spatial distribution follows the polarization state of the laser beam incident on the unpolarized relativistic electron beam~\cite{Sun2011_STAB,petrillo2015,zhijun_chi2020,Hajima2021,paterno_2022}. For linearly polarized laser photons, the LCS $\gamma$-ray beam has an ellipse-like spatial intensity profile, with its major axis perpendicular to the direction of the polarization. For unpolarized and circularly polarized laser beams, the spatial intensity distribution is azimuthally symmetric. Such LCS $\gamma$-ray beam spatial distribution measurements have been reproduced by several dedicated simulation codes, such as the \textsc{mccmpt}~\cite{Sun2011_STAB} validated on HI$\gamma$S~\cite{litvinenko_1997} experimental data, the \textsc{mclcss} code~\cite{Luo2011} and the code described in ref.~\cite{Brown_Hatermann_2004} validated on PLEIADES experimental data~\cite{Brown_2004}, the \textsc{cmcc} code~\cite{Curatolo_PhD_thesis} applied in ref.~\cite{Curatolo2017} for characterization of the STAR-I, EuroGammaS and XFELO $\gamma$-sources.  

However, the spatial and energy distribution of the LCS $\gamma$-ray beam impinged on the target is controlled through a collimation system, which selects the backscattered highest energy component of the Compton spectrum and conveniently limits the beam spot size. Circular aperture collimators have been used at the HI$\gamma$S~\cite{CSun_PhD_thesis}, TERAS~\cite{Ohgaki1991,Ohgaki2000} and NewSUBARU~\cite{amano09,Horikawa2010} LCS $\gamma$-ray sources. The use of rectangular profile collimators in connection with electron beams with unequal transverse emittances, characteristic for electron storage rings, has been recently proposed in ref.~\cite{Hajima2021}. A double collimation system designed for shaping regular dodecagonal beam spots at the target position~\cite{hao_fan_2022} is installed at the SLEGS~\cite{wang_fan_2022} LCS $\gamma$-ray source of the Shanghai synchrotron radiation facility~\cite{Cao2022}. 

In contrast with the characterization of LCS $\gamma$-ray beams spatial distribution in terms of collimation settings, the effect of the electron and laser beam parameters on the collimated LCS $\gamma$-ray beams spatial distribution has sparsely been investigated. In this work, we make a systematic investigation of the spatial distribution of collimated LCS $\gamma$-ray beams as a function of electron beam energy and emittance, laser wavelength, focusing conditions and polarization, as well as the bremsstrahlung background radiations generated by the electron beam. For the investigation, we use the \texttt{eliLaBr}~\cite{Filipescu_2022_POL,Filipescu_2023_LCS,eliLaBr_github} Monte Carlo modeling code and experimental LCS $\gamma$-ray beam spatial distributions measured at the NewSUBARU facility.  

\section{NewSUBARU $\gamma$-ray beam line}
\label{sec_NS}

\begin{figure*}[h!]
\centering
\includegraphics[width=0.85\textwidth, angle=0]{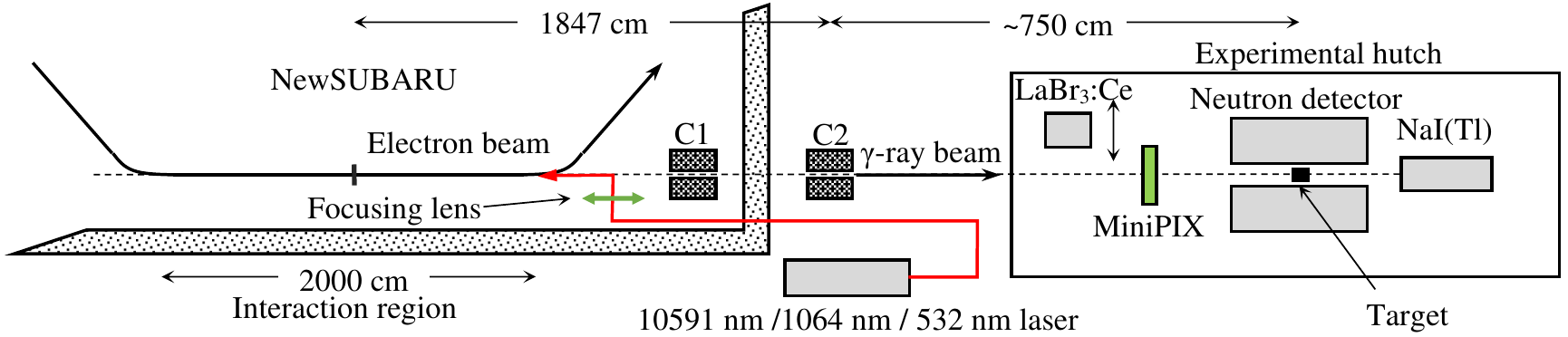}  
\caption{The $\gamma$-ray beam line of the NewSUBARU synchrotron radiation facility.}
\label{fig-bl01}       % Give a unique label
\end{figure*}

Laser photons are sent head-on against relativistic electrons in the BL01 beamline of NewSUBARU synchrotron radiation facility and Compton scattered $\gamma$-beams are produced. The LCS $\gamma$-ray beams pass through a collimator system thus providing a narrow bandwidth. The layout of the NewSUBARU LCS $\gamma$-ray beamline BL01 and of the experimental setup is shown in figure~\ref{fig-bl01}. 

\begin{table}[htbp]
\centering
\caption{\label{tab_lasers} Parameters for the CO$_2$ laser and the solid state INAZUMA and Talon lasers used at the NewSUBARU BL01 LCS $\gamma$-ray beam line: wavelength ($\lambda$), active medium, pumping type, pulse frequency ($f_p$), full angle divergence ($\theta_l$) and diameter at laser output (D$_{4\sigma}$).}
\smallskip
\begin{tabular}{|l|r|c|l|r|c|c|}
\hline
Laser   &$\lambda$         &active        &pumping type  &$f_p$  &$\theta_l$ &D$_{4\sigma}$\\
        & [$\mu$m]         &medium        &              &[kHz]  & [mrad] &[mm]\\
\hline
CO$_2$  & 10.591           & CO$_2$       & electrical gas discharge    & 10-25   & $<$7.6   & 1.75\\
INAZUMA & 1.064            & Nd:YVO$_4$   & diode-pumped                & 10-25 & $<$2.5   & 0.6\\
Talon   & 0.532            & undisclosed  & diode-pumped                & 1-25  & $<$0.9   & 1.048\\
\hline
\end{tabular}
\end{table}

Electron beams with 982 MeV energy and typical currents of $\sim$300~mA are injected in the NewSUBARU ring. Starting from the injection energy, the electron beam energy can be continuously varied either in deceleration mode, down to 0.5~GeV, or in acceleration mode, up to 1.5~GeV. A 10.591~$\mu$m wavelength CO$_2$ laser and two diode-pumped solid state lasers, the 1.064~$\mu$m Nd:YVO$_4$ INAZUMA laser and the 0.532~$\mu$m Talon laser, are routinely used at NewSUBARU. Table~\ref{tab_lasers} gives the main parameters of the three lasers. Thus, LCS $\gamma$-ray beams with maximum energies between 0.5~MeV and 76~MeV can be produced, following the well known expression for the $E_\gamma$ energy corresponding to Compton backscattered photons in head-on collisions:
\begin{equation}\label{eq_eg_energy}
E_\gamma=\cfrac{4\gamma^2E_p}{1+(\gamma\theta)^2+4\gamma E_p/(mc^2)}.
\end{equation}
Here $E_p$ is the laser photon energy, $mc^2$ is the electron rest mass, $\theta$ is the scattering angle of the $\gamma$-ray photon relative to the electron incident direction, and $\gamma$ is the Lorentz factor of the incident electron. 

A system of two 10 cm thick lead collimators is used to limit the $\gamma$-ray beamspot size and define the energy resolution. The two collimators denoted C1 and C2 are placed at 15.47~m (inside the accelerator radiation shield) and respectively at 18.47~m downstream from the center of the straight electron beamline BL01. The collimators have circular apertures which can take values between 6~mm and 1~mm. They are mounted on $x$-$y$-$\theta$ stages driven by stepping motors which allow fine tunings of their alignment on the horizontal ($x$), vertical ($y$) and rotational axes ($\theta$). The collimated LCS $\gamma$-ray beam irradiates the targets placed at the center of a moderated neutron detection array placed in the experimental hutch, as shown in figure~\ref{fig-bl01}. 

\begin{figure*}[h]
\centering
\includegraphics[width=0.70\textwidth, angle=0]{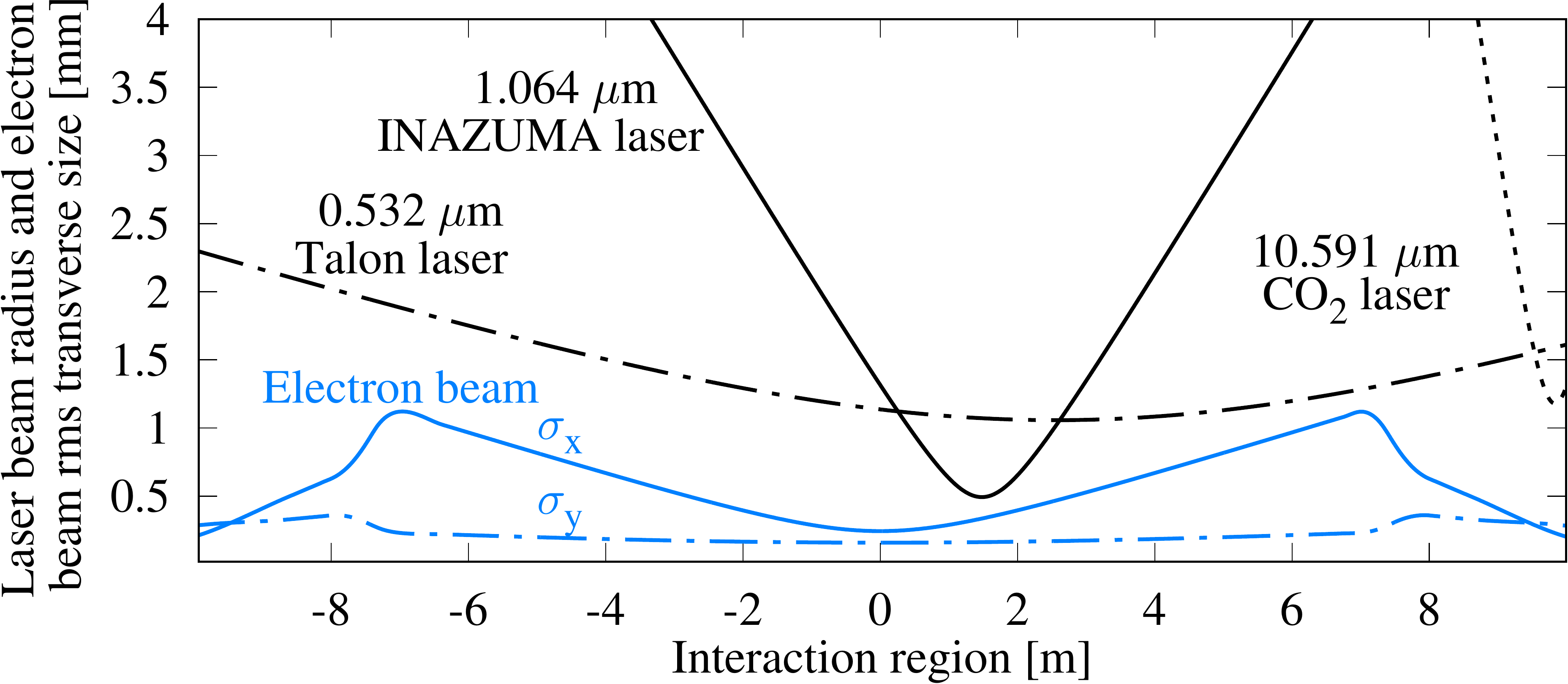}  
\caption{Horizontal  ($\sigma_x$ solid blue line) and vertical ($\sigma_y$ dashed dotted blue line) electron beam rms transverse sizes obtained using the MAD-X beam parameters shown in figure 5 of ref.~\cite{Filipescu_2023_LCS} and considering the nominal emittance of ($\varepsilon_x$,~ $\varepsilon_y$)~=~(38,~3.8)~nm$\cdot$rad. 
Gaussian optics calculations of the transverse beam radii for the INAZUMA laser of $\lambda$~=~1.064~$\mu$m (solid black line), Talon laser of $\lambda$~=~0.532~$\mu$m (dashed dotted black line) and CO$_2$ laser of $\lambda$~=~10.591~$\mu$m (dotted black line).}
\label{fig_LCS_overlap_2}       % Give a unique label
\end{figure*}

\subsection{Time and spatial distributions of the electron and laser beams}
\label{sub_sec_spatial_distrib_laser_electron}

At NewSUBARU, the electron beam spatial profile is characterized by Gaussian distributions along the transverse horizontal ($x$) and vertical ($y$) axes. The nominal electron beam emittance is of ($\varepsilon_x$,~ $\varepsilon_y$)~=~(38,~3.8)~nm$\cdot$rad and the rms energy spread at the injection energy of 982~MeV is 0.04$\%$. Figure~\ref{fig_LCS_overlap_2} shows the transverse electron beam size on horizontal (solid blue line) and vertical (dashed dotted blue line) axes in the interaction region, obtained with the MAD-X calculations given in figure 5 of ref.~\cite{Filipescu_2023_LCS}. The laser beam transverse intensity profile is characterized by Gaussian functions in the horizontal and vertical axes with equal standard deviation. Figure~\ref{fig_LCS_overlap_2} shows the CO$_2$, INAZUMA and Talon laser beams radii along the interaction region, where the beam radius $w_u$ is defined accordingly to the $D4\sigma$ definition of the laser diameter, as two standard deviations of the intensity distribution $I(x,y)$:
\begin{equation}
w_u \equiv 2 \sqrt{\cfrac{\int_x\int_y u^2\cdot I(x,y)dxdy}{\int_x\int_y I(x,y)dxdy}}.
\end{equation}
Here $u$ stands for either the horizontal $x$ or vertical $y$ coordinates, which are taken to be relative to the beam center. The lasers used in the present study have equal radii on the two perpendicular transverse directions. More details on the electron and laser beams spatial distributions can be found in references~\cite{amano09,Horikawa2010,Utsunomiya2015,Filipescu_2023_LCS}. 

The NewSUBARU electron beam bunches have a 500~MHz frequency and 60~ps width. The lasers are typically operated in Q-switch mode at much lower frequencies, between 1 and 25~kHz, and have much wider photon pulses of tens of ns. Thus, the LCS $\gamma$-ray beam time structure follows the slow laser and the fast electron time structure. Neutron emission spectra measurements with the time-of-flight method make use of synchronized laser and single-bunch electron beams~\cite{miyamoto_2018,kirihara_2020}, where only one electron bunch circulates in the ring and collides with the laser pulse at a designed position. The electron beam current for the single-bunch operation mode is limited to $\sim$20~--~30~mA. Photoneutron cross section measurements with a moderated neutron detection array are generally operated in higher $\gamma$-ray flux conditions, using unsynchronized laser and multiple-bunch electron beams. These are the usual operating conditions at NewSUBARU, for which the maximum stored current is 300~mA. In this mode, the Compton interactions can take place along the entire straight beamline section, according to the laser beam-electron beam spatial overlap along the beamline. The laser-electron overlap function for the three lasers operated at NewSUBARU has been recently discussed in ref.~\cite{Filipescu_2023_LCS}.  

\subsection{LCS $\gamma$-ray beam parameters and beam diagnostics methods at NewSUBARU}
\label{sub_sec_beam_diagnostics}

From equation~\ref{eq_eg_energy}, it follows that the energy of the electron beam and that of the laser photon directly determine the maximum energy of the LCS $\gamma$-ray beam, for $\theta$~=~0. At NewSUBARU, the absolute $\gamma$-ray beam energy is known with a precision of the order of 10$^{-5}$ due to the energy calibration of the electron beam \cite{IEEE_Utsunomiya14,Shima14}, which was performed using low-energy LCS $\gamma$-ray beams produced with a grating-fixed CO$_2$ laser.

Typical collimated $\gamma$-ray beam flux values are of $\sim$10$^5$ $\gamma$/s for 25~kHz laser operation and $\sim$10$^4$ $\gamma$/s for 1~kHz laser operation. The low frequency, 1~kHz operation mode is necessary for neutron-multiplicity sorting experiments with moderated neutron detection arrays~\cite{utsunomiyaNimDNM,IGheorghe_2021_MF}, where the time interval between consecutive photon pulses must be longer than the neutron moderation time in the detector. The absolute flux of the incident $\gamma$-ray beam is monitored with a NaI(Tl) detector placed downstream of the neutron detection system, as shown in figure~\ref{fig-bl01}, and the associated pile-up or Poisson fitting method~\cite{kondo11,utsunomiya_nimMP_2018} of unfolding the recorded multiple-photon summed spectra.

The energy spread of the NewSUBARU LCS $\gamma$-ray beams varies between 1$\%$ and 3$\%$ in full width at half maximum (FWHM) in good electron beam conditions and it is measured with a large volume LaBr$_3$:Ce detector. The incident photon spectrum is obtained by reproducing the experimental detector response using the dedicated \texttt{eliLaBr} Geant4 code described in section~\ref{sec_elilabr}. 

\section{\texttt{eliLaBr} LCS $\gamma$-ray source simulation code}
\label{sec_elilabr}
 
The interpretation of the experimental results obtained in the IAEA CRP series of GDR photoneutron cross section measurements performed at NewSUBARU required the description of the incident LCS $\gamma$-beams spectral distribution. As mentioned in the introduction, there are several Monte Carlo simulation codes described in literature~\cite{Sun2011_STAB,Curatolo2017,Hajima2021,paterno_2022}. The existing LCS simulation codes can treat incident beams which collide so that the centers of the laser and electron beam pulses overlap at the designed interaction point, without any position and timing jitters. However, at NewSUBARU, the laser beam and the electron beam sent head-on against each other are generally unsynchronized. Thus, the interactions between the laser pulses and the electron pulses can take place along the entire straight electron beamline BL01, following the spatial overlap between the laser beam and the electron beam along the interaction region. Therefore, it was necessary to develop a method to model the Compton interaction between continuous, unsynchronized laser and electron beams sent head-on against each other, with realistic treatment of the spatial overlap between the laser and electron beams along the interaction region. The method has been implemented in the \texttt{eliLaBr} simulation code.

The \texttt{eliLaBr} spectral distribution and flux results have been validated in ref.~\cite{Filipescu_2023_LCS} against experimental detector responses of NewSUBARU LCS $\gamma$-ray beams produced with 982~MeV electrons and 10.591~$\mu$m, 1.064~$\mu$m and 0.532~$\mu$m wavelength laser beams. Here, we briefly describe the main features of the code and guide the reader to refs.~\cite{Filipescu_2023_LCS,Filipescu_2022_POL} for more details.

\subsection{Laser and electron beam modeling} 

The central energy, the energy spread and the transverse emittance values $\varepsilon_{x,y}$ of the electron beam are treated as input parameters in the simulation code. 
The electron beam spatial profile is characterized by normal distributions along the transverse axes, with the $dN_e(x,y,z)$ electron density given by~\cite{Horikawa2010}:
\begin{equation}\label{label_eq_dNe}
dN_e(x,y,z) = \cfrac{I}{ev} \cdot dn_e(x,y,z),
\end{equation}
where $I$ is the electron beam current, $e$ and $v$ are the electron charge and velocity, and
\begin{equation}\label{label_eq_dNe}
dn_e(x,y,z) = \cfrac{1}{2\pi\sigma_x(z)\sigma_y(z)}\exp\left[-\cfrac{1}{2}\left(\cfrac{x}{\sigma_x(z)}\right)^2-\cfrac{1}{2}\left(\cfrac{y}{\sigma_y(z)}\right)^2\right].
\end{equation}
Here, the standard deviations $\sigma_{x,y}(z)$ in the horizontal and vertical axes along the interaction region are described by the emittance, betatron functions and the horizontal dispersion functions. The electron beam phase space distribution along the interaction region is determined using the Twiss parameters $\alpha$ and $\beta$, the $\psi$ betatron phase advance and $D$, $D'$ dispersion functions along the interaction beam line, which are either read in tabulated format from an input file, or internally calculated assuming a drift section along the entire beam line. 

The laser beam wavelength, energy spread, polarization state, diameter, quality factor $M^2$, as well as optical parameters of the focusing lens are treated as input paratemers in the simulation code. The laser beam transverse intensity profile is described by normal distributions with equal standard deviations in the horizontal and vertical axes, with laser photon density $dN_p(x_p,y_p,z_p)$ given by:
\begin{equation}\label{label_eq_dNp}
dN_p(x_p,y_p,z_p)=\cfrac{P}{E_p c} \cdot dn_p(x_p,y_p,z_p),
\end{equation}
where $P$ is the laser power, $E_p$ is the energy of the laser photon, $c$ is the speed of light, and
\begin{equation}\label{label_eq_dNp_overlap}
dn_p(x_p,y_p,z_p)=\cfrac{1}{2\pi\sigma_p^2}\exp\left(-\cfrac{1}{2}\cfrac{x_p^2+y_p^2}{\sigma_p^2} \right).
\end{equation}
The standard deviation $\sigma_p(z_p)$ along the laser beam path is obtained using Gaussian optics calculations. We note that we have referred to the position in the electron beam coordinate system as $(x,y,z)$, and as $(x_p,y_p,z_p)$ in the laser beam one. 

The code also takes into account:
\begin{itemize}
\item non-zero longitudinal $\Delta z$ displacements between the electron beam focus and the laser beam focus determined by the laser conditions ($\lambda$, beam radius at laser output, qualify factor M$^2$) and the optics system;  
\item non-zero transverse offsets $\Delta x$ and $\Delta y$ between the electron beam axis and the laser beam axis;
\item small $\theta$ and $\varphi$ rotations between the laser and electron beam axes. 
\end{itemize}
At NewSUBARU, the laser focus position is conveniently chosen to maximize the laser-electron luminosity, which generally leads to non-zero $\Delta z$ displacement values. Non-ideal alignment settings can generate non-zero transverse $\Delta x$ and $\Delta y$ off-sets and laser-electron beam axes rotations. Small misalignments can significantly reduce the LCS $\gamma$-ray beam flux and deteriorate its spectral distribution. The alignment precision as a function of the laser polarization has been discussed in ref.~\cite{Filipescu_2023_LCS}.  

\subsection{$\gamma$-ray beam flux} 

The $\mathcal{L}$ luminosity of the LCS $\gamma$ ray beam is also computed in terms of the $\texttt{L}$($z$) laser beam-electron beam overlap function, as:
\begin{equation}\label{label_eq_luminosity_aligned}
\mathcal{L} = c(1+\beta\cdot \cos\theta)\cdot\cfrac{I}{ev}\cfrac{P}{E_p c} \int_z \texttt{L}(z) dz,
\end{equation}
where the $\cos \theta$ term takes into account the angle between the laser and the electron beams and $\texttt{L}$($z$) is defined as:
\begin{equation}\label{eq_overla_function}
\texttt{L}(z) = \int_x \int_y dn_e(x,y,z) dn_p(x,y,z) dx dy.
\end{equation}
The number of Compton scattered $\gamma$-rays per unit time is futher expressed as~\cite{Sun2011_STAB}:
\begin{equation}\label{label_eq_dNg_dt}
\cfrac{dN_\gamma}{dt} = \mathcal{L}\cdot\sigma_{tot},
\end{equation}
where $\sigma_{tot}$ is the angle integrated cross section for the Compton scattering of laser photons on relativistic electrons for head-on collisions.

\subsection{Sampling procedure} 

The Compton interaction position along the straight section of the electron beam line is sampled accordingly to the $\texttt{L}$($z$) laser beam and electron beam overlap function. The interaction point transverse electron phase-space coordinates ($x$,~$y$,~$dx/dz$,~$dy/dz$) are sampled in the electron beam focus following the procedure described in ref.~\cite{Sun2011_STAB} (Eqs.~56). Here we consider a Gaussian electron phase-space distribution and the electron beam focus placed in a drift section of the electron beam line. The $\mathbf{M}$ transfer matrix for particle trajectories~\cite{HelmutWiedemann_PAP} is then used to compute the electron phase-space coordinates at the $z$ sampled position along the beam line. The sampled interaction point is either accepted or rejected based on the laser photon density $dN_p(x_p,y_p,z_p)$ given in eq.~\eqref{label_eq_dNp}. If accepted, electron and laser quadrivectors are generated in the sampled interaction point. The energy of the electron and laser photon are sampled accordingly to their corresponding energy spread values. The laser photon momentum is obtained from the gradient of the laser propagation phase for a Gaussian beam~\cite{Sun2011_STAB}.  

The sampled electron and laser photon momentum quadrivectors are rotated and Lorentz transformed to obtain a head-on collision in the electron rest frame system, following the well known procedure used in the existing Monte Carlo LCS simulation codes~\cite{CAIN,Sun2011_STAB,Curatolo2017,Curatolo_PhD_thesis}. The polarization of the scattered photon is treated independently both in the Stokes parameters and in the polarization vector formalisms, and it is fully implemented in the Monte Carlo method. Thus, the transformations applied to the laser momentum quadrivector are also applied to the laser polarization quadrivector, followed by a gauge invariance transformation and the extraction of the Stokes polarization vector~\cite{McMaster1,McMaster2}. The code provides the polarization properties of the scattered photon in the particle reference system, according to the specific requirements of the \textsc{Geant4} framework. The \texttt{eliLaBr} results of LCS $\gamma$-ray photon beam polarization spatial distributions are validated in ref.~\cite{Filipescu_2022_POL} against analytical results given in refs.~\cite{Sun2011_STAB,petrillo2015,zhijun_chi2020,zhijun_chi2022}.  

\subsection{Open source implementation into end-to-end experiment simulation tool} 
The Monte Carlo LCS modeling procedure was implemented in a \texttt{C++} class integrated into the \texttt{eliLaBr} simulation tool for photonuclear experiments. The code runs under \textsc{Geant4.9}~\cite{geant_agostinelli_2003,geant_allison_2006,geant_allison_2016} and it is available on the GitHub repository~\cite{eliLaBr_github}. An updated version compatible with \textsc{Geant4.10} and \textsc{Geant4.11} is soon to be uploaded on GitHub. 

The geometry of the experimental setup implemented in the \texttt{eliLaBr} code reproduces the LCS $\gamma$-ray beamline BL01 present at NewSUBARU:
\begin{itemize}
\item Incident LCS $\gamma$-ray flux and energy distribution monitors;
\item A moderated 4$\pi$ neutron detection system of flexible design;    
\item The primary (C1) and secondary (C2) $\gamma$-beam collimators with variable apertures, as well as a large aperture $\gamma$-beam shutter and concrete walls;
\item Synchrotron dipole magnetic field;
\item Vacuum beam pipe, laser optical mirror used to insert the laser beam into the electron beamline and borosilicate glass vacuum window;
\item Variable vacuum conditions inside the electron accelerator and air at atmospheric pressure in the experimental hall.
\end{itemize}

\section{Simulation of collimated LCS $\gamma$-ray beams -- transverse spatial and energy distributions}
\label{sec_sp_distrib}

\begin{figure*}[h!]  % \columnwidth
\centering
\includegraphics[width=0.47\textwidth, angle=0]{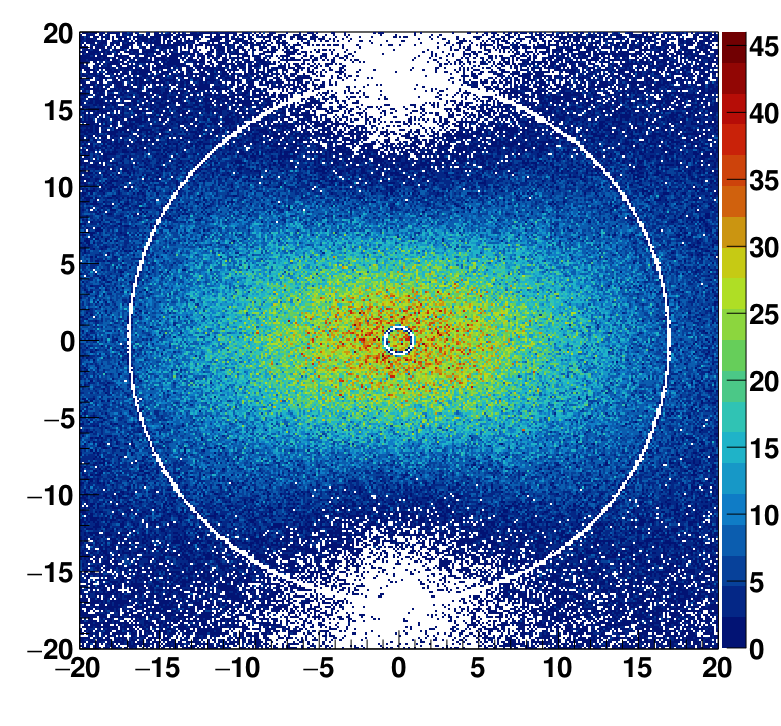} \quad
\includegraphics[width=0.47\textwidth, angle=0]{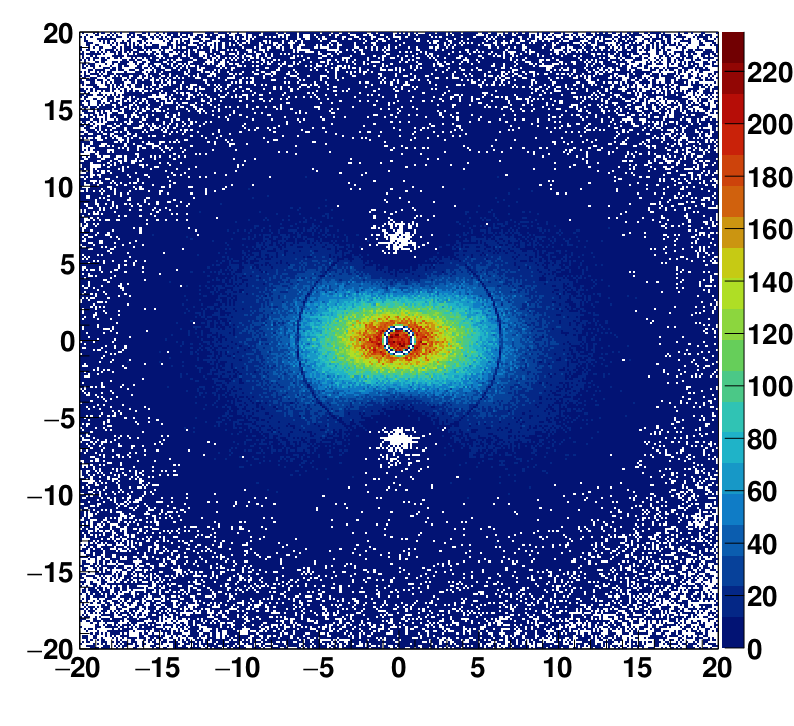}
\put(-420,80){\rotatebox{90}{Vertical [mm]}}
\put(-205,80){\rotatebox{90}{Vertical [mm]}}
\put(-330,0){Horizontal [mm]}
\put(-125,0){Horizontal [mm]}
\put(-425,172){\large{(a)}}
\put(-210,172){\large{(b)}}
\caption{Simulated spatial distributions of LCS $\gamma$-ray beams produced with the INAZUMA laser ($\lambda$~=~1.064~$\mu$m) and (a) 559~MeV and (b) 1480~MeV electron beams at the position of the C2 collimator, 18.47~m downstream from the center of the electron beamline. Ideal pencil-like electron and laser beams interacting in the electron beam focus are considered. The simulations were performed by removing the upstream collimator C1. The inner white ring represents the C2 collimator opening, of 2~mm diameter. The outer (a) white and (b) blue rings show the $1/\gamma$ emission angle. 100$\%$ vertically polarized laser beams were considered.     
}
\label{fig_inv_gamma}       % Give a unique label
\end{figure*}
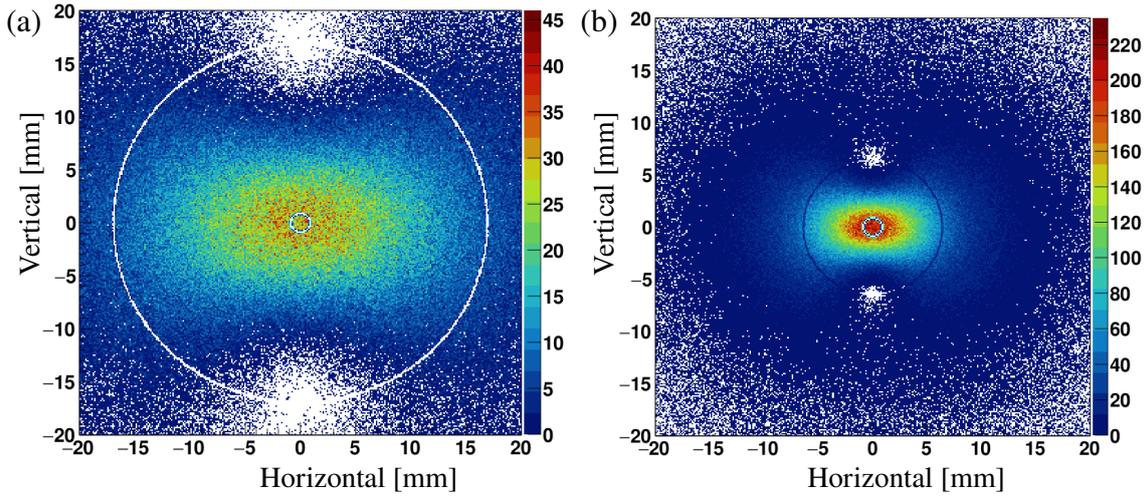

A double collimation system of C1~=~3~mm and C2~=~2~mm apertures is generally used at NewSUBARU for photoneutron experiments, for which the corresponding flux and energy spread values have been discussed in section~\ref{sub_sec_beam_diagnostics}. Figure~\ref{fig_inv_gamma} shows simulated spatial distributions of LCS $\gamma$-ray beams produced with the 100$\%$ vertically polarized INAZUMA laser and (a) 559~MeV and (b) 1480~MeV electron beams at the position of the C2 collimator. The simulations were performed by removing the upstream C1 collimator. The results reproduce the well-known spatial distribution of uncollimated LCS $\gamma$-ray beams produced with relativistic electrons and 100$\%$ linearly polarized laser photons~\cite{Sun2011_STAB,petrillo2015,zhijun_chi2020,Hajima2021,Filipescu_2022_POL}. The photons are mostly scattered in the horizontal plane, perpendicular to the laser polarization vector.

The inner rings of 2~mm diameter present in figure~\ref{fig_inv_gamma} show the opening of the C2 collimator. The outer rings mark the limits of the cone with half-opening angle of $1/\gamma$, where $\gamma$ is the Lorentz factor of the incident electron. About half of the $\gamma$-rays produced through Compton scattering of laser photons on relativistic electrons are scattered into the narrow cone with half-opening of $1/\gamma$~\cite{Sun2011_STAB}. We notice in figure~\ref{fig_inv_gamma} that the collimator cuts well within the $1/\gamma$ cone and the spatial distribution of the collimated LCS $\gamma$-ray beam is unaffected by the anisotropy given by the laser polarization orientation. However, the distributions shown in figure~\ref{fig_inv_gamma} have been obtained using ideal pencil-like laser and electron beams interacting in a well defined point (at the electron beam focus). When considering realistic laser and electron beams interacting along the entire electron beamline, the spatial distribution anisotropy correlated to the laser polarization smears out. Also, the 1/$\gamma$ rings are no longer well defined, as their radius values depend on the interaction point position along the beamline.

Considering different laser optics and laser polarization conditions, as well as different electron beam energies and phase space distributions, we investigate in the following subsections the intensity spatial distributions and energy spatial distributions on an imaging plate. The results are obtained using the \texttt{eliLaBr} simulation code for the C1~=~3~mm and C2~=~2~mm collimators configuration. The virtual 14~$\times$~14~mm$^2$ imaging plate is placed at 25.5~m downstream of the NewSUBARU BL01 beamline midpoint, respectively at 7~m downstream of the second (C2) collimator. This corresponds to the target position in photoneutron experiments. We considered LCS $\gamma$-ray beams produced at NewSUBARU with the INAZUMA laser of $\lambda$~=~1.064~$\mu$m and electron beam energies in the 0.5~--~1.5~GeV range. 

\subsection{Laser optics influence}
\label{sub_sec_laser_mirror}

\begin{figure*}[h!]  % \columnwidth
\centering
\includegraphics[width=0.95\textwidth, angle=0]{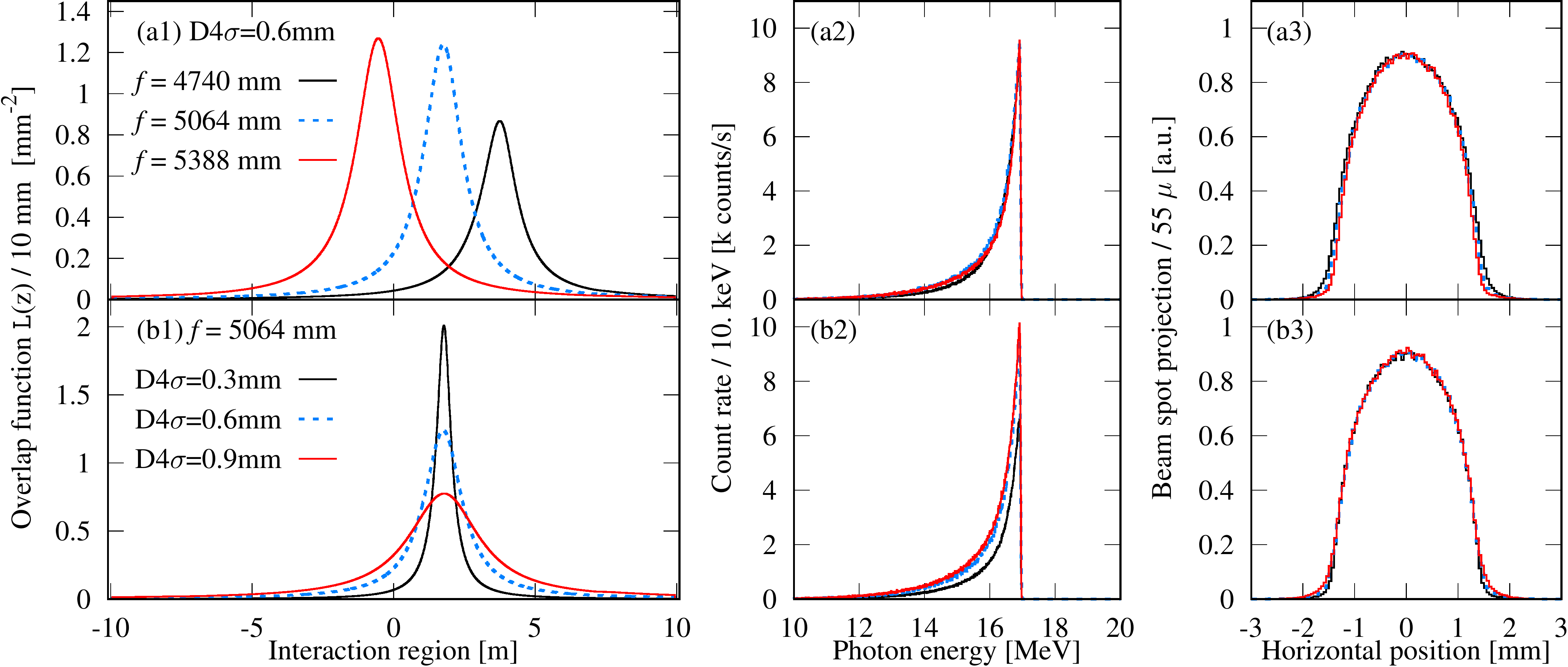} % {ETC/FIG_LFD_0974/fig_LFD_0974.pdf} 
\caption{Simulations of LCS $\gamma$-ray beam spectra and spatial distribution for $E_e$~=~982~MeV electron beams with nominal emittance ($\varepsilon_x$,~$\varepsilon_y$)~=~(38,~3.8)~nm$\cdot$rad, and different configurations of the $\lambda$~=~1064~nm INAZUMA laser: (a1--3) study for different focal lengths of the focusing lens, where the nominal $f$~=~5064~mm; (b1--3) study for different laser beam diameter values D4$\sigma$ at laser output, where the nominal D4$\sigma$~=~0.6~mm. Left column: laser-electron beams overlap function along the beamline. Middle column: count rate for collimated incident photon spectra. Right column: horizontal projection of the LCS $\gamma$-ray beam spatial distribution incident on target.     
}
\label{fig_LFD_0974}       % Give a unique label
\end{figure*}

Figure~\ref{fig_LFD_0974} shows simulations of collimated LCS $\gamma$-ray beam spectra and spatial distributions on target for different optics conditions of the INAZUMA laser and fixed electron beam parameters: ($\varepsilon_x$,~$\varepsilon_y$)~=~(38,~3.8)~nm$\cdot$rad emittance, which is considered good, and injection energy of 982~MeV. This corresponds to $E_\gamma$~=~16.95~MeV maximum energy LCS $\gamma$-ray beams. The left column shows the laser-electron beam overlap function \texttt{L}($z$) along the beamline, where \texttt{L}($z$) is defined in equation~\eqref{eq_overla_function}. The second column in figure~\ref{fig_LFD_0974} shows the collimated incident spectra. The horizontal projections of the LCS $\gamma$-ray beam spatial distributions incident on target are shown in the right column.  

In the upper row of figure~\ref{fig_LFD_0974}, we consider different focal length values for the focusing lens. We have shown in figure~6(b) of ref.~\cite{Filipescu_2023_LCS} that, for low-divergence, short-wavelength laser beams such as the ones emitted by the 532~nm Talon laser, the laser-electron interactions take place along the entire length of the straight beamline section. In contrast, figure~\ref{fig_LFD_0974}(a1) shows that for the higher divergence 1064~nm INAZUMA laser, the laser-electron interaction region is limited to $\sim$6~m and the beam overlap function resembles a Gaussian function. The focal length of the lens defines the Gaussian centroid position, and thus the laser-electron interaction region for the INAZUMA laser case: a smaller focal length brings the interaction region closer to the collimator. Figure~\ref{fig_LFD_0974}(a3) shows that the beam spot increases with decreasing the focal length, as expected. However, figure~\ref{fig_LFD_0974}(a2) shows that the spectrum does not deteriorate with decreasing the distance between interaction point and collimator. This is explained by the different electron beam phase space distributions probed along the three different interaction regions.     

For the low-divergence 532~nm Talon laser, we have shown in figure~7(a1) of ref.~\cite{Filipescu_2023_LCS} that the distance between the interaction region centroid and the collimator increases with decreasing the D4$\sigma$ diameter at laser output. In contrast, figure~\ref{fig_LFD_0974}(b1) shows that, for the higher divergence 1064~nm INAZUMA laser, the centroid of the laser-electron overlap function is insensitive to the laser diameter at laser output, although the length of the effective interaction region, and thus the LCS $\gamma$-ray beam flux shown in figure~\ref{fig_LFD_0974}(b2), are strongly affected by it. However, the beam spot size on the imaging plate is only slightly affected to strong variations in the laser diameter at laser output, as shown in figure~\ref{fig_LFD_0974}(b3).  

\subsection{Laser polarization influence}
\label{sub_sec_laser_pol}

\begin{figure}[h!]  % \columnwidth
\centering
\includegraphics[width=0.9\textwidth, angle=0]{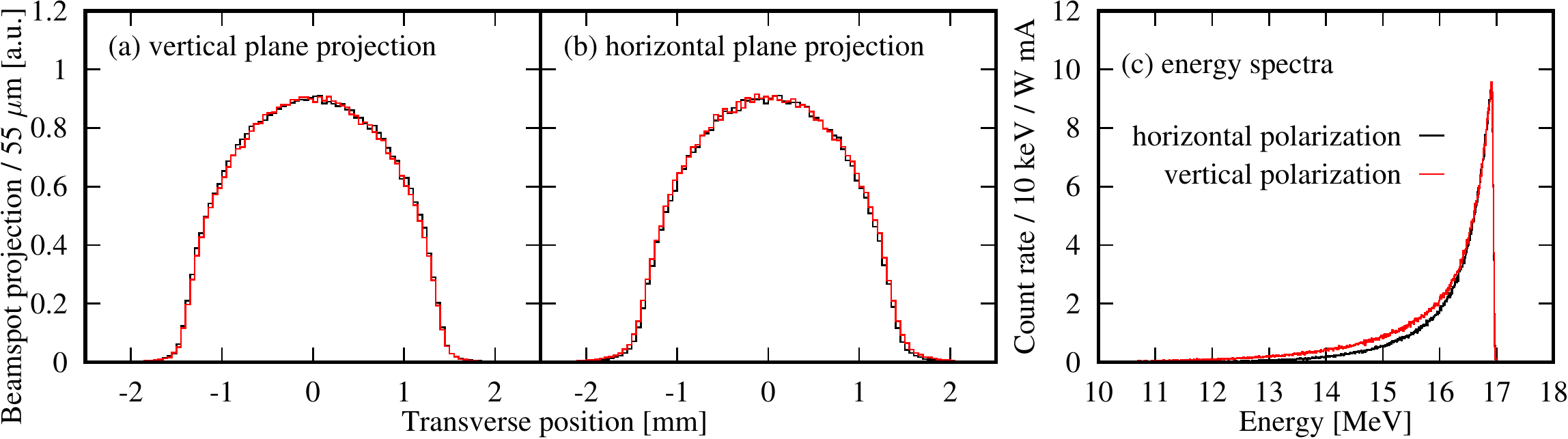} %{ETC/FIG_POL/fig_pol.pdf} 
\caption{Simulations of LCS $\gamma$-ray beam spatial distribution and spectra for $E_e$~=~982~MeV and different polarization configurations of the $\lambda$~=~1064~nm INAZUMA laser. (a) Vertical and (b) horizontal beamspot projections for lasers with 100$\%$ horizontal ~--~in the accelerator plane (black lines) and vertical (red lines) linear polarizations. The corresponding incident energy spectra are shown in (c).
}
\label{fig_pol}       % Give a unique label
\end{figure}

We have recently investigated~\cite{Filipescu_2023_LCS} the influence of laser beam polarization plane orientation on the energy spectrum of collimated LCS $\gamma$-ray beams produced using electron beams of unequal transverse emmitance profiles characteristic to electron storage rings. We have shown that the use of laser beams linearly polarized along the horizontal plane improves the LCS $\gamma$-ray beam energy resolution. Figure~\ref{fig_pol} shows that, although the spectral distribution is affected by the laser polarization, the beam spot shape is insensitive to the laser polarization plane. 

\begin{figure}[h!]  % \columnwidth
\centering
\includegraphics[width=0.99\textwidth, angle=0]{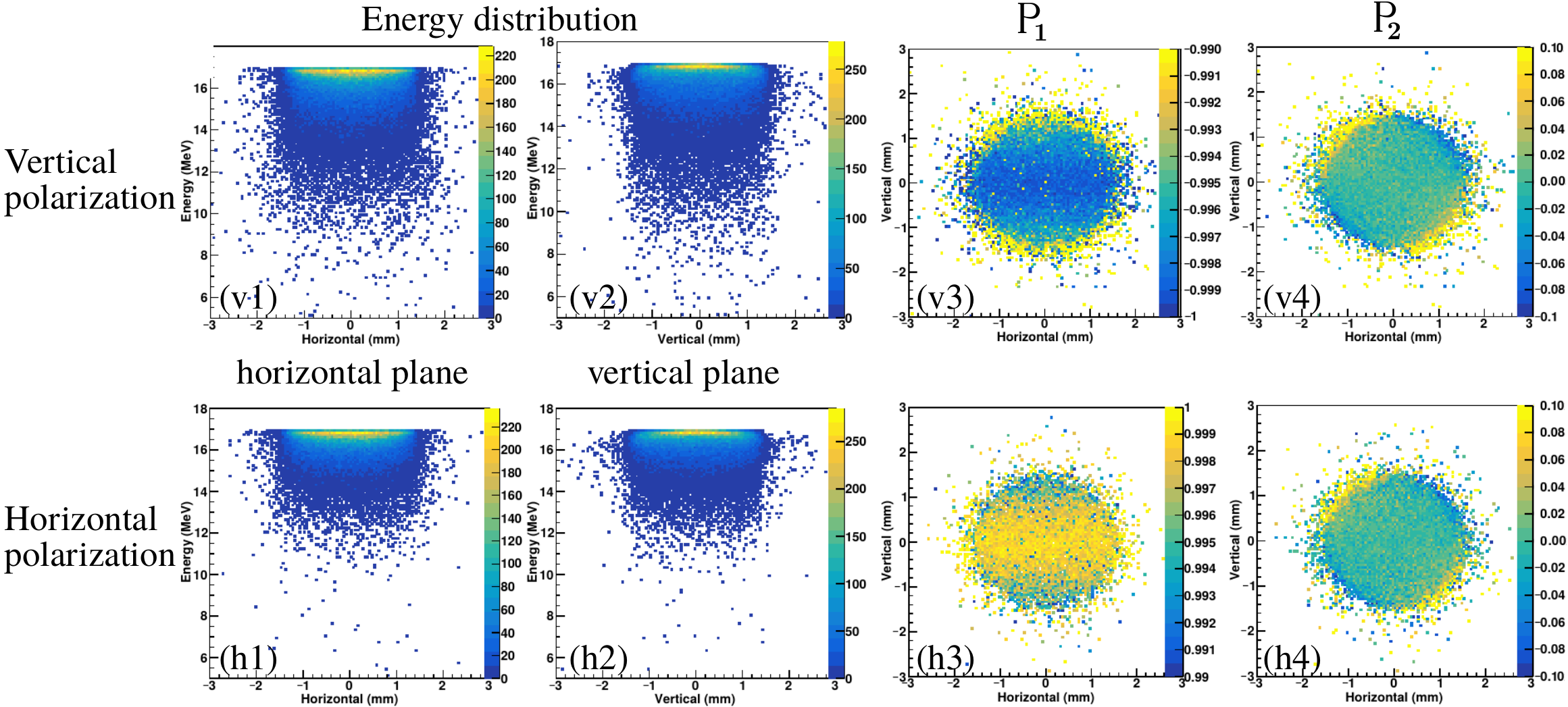} %{ETC/FIG_POL/fig_pol_maps.png} 
\caption{Simulations of 16.95~MeV maximum energy LCS $\gamma$-ray beams produced with the $\lambda$~=~1064~nm INAZUMA laser and 982~MeV electron beams of good ($\varepsilon_x$,~$\varepsilon_y$)~=~(38,~3.8)~nm$\cdot$rad emittance. Laser beams 100$\%$ linearly polarized in the (v1--4) vertical and (h1--4) horizontal planes have been considered. The energy spatial distributions at the imaging plate position in the (v1, h1) horizontal and (v2, h2) vertical planes. The (v3, h3) $P_1$ and (v4,h4) $P_2$ Stokes parameters spatial distributions.}
\label{fig_pol_maps}       % Give a unique label
\end{figure}

Figures~\ref{fig_pol_maps} show the energy spatial distributions in the (v2,h2) vertical and (v1,1) horizontal planes obtained using 100$\%$ (v1,v2) vertically and (h1,h2) horizontally polarized lasers. We notice the improved energy resolution of LCS $\gamma$-ray beams produced with horizontally polarized lasers. 

Figures~\ref{fig_pol_maps}(v1,v2) and (h1,h2) show that there is an asymmetry in the transverse spatial distribution of maximum energy photons, $\sim$17~MeV in this case. It can be observed that photons of maximum energy have a wider spatial distribution on the horizontal axis than on the vertical one. The effect is present both when using 100$\%$ vertically and horizontally polarized lasers. Therefore, not being correlated with the polarization direction of the laser, we can deduce that the transverse asymmetry of the spatial distribution of LCS $\gamma$-rays can only be determined by the transverse asymmetry of the phase space distribution of electrons. A factor of ten between the vertical and the horizontal emittance values, characteristic to the NewSUBARU synchrotron, has been considered in all the simulations presented in this work.

Figures~\ref{fig_pol_maps}(v3,4) and (h3,4) show the corresponding spatial distributions of the Stokes $P_1$ and $P_2$ polarization parameters for the scattered LCS $\gamma$-ray beams, where the parameter $P_1$ describes the linear polarization relative to a coordinate system defined by the vertical and horizontal axes of the accelerator ring, and $P_2$ the linear polarization relative to a coordinate system rotated by an angle of $\pi$/4 to the right. We notice the well known ($P_1$,~$P_2$) pairs of (-1,~0) for vertical polarized laser and (+1,~0) for horizontal polarized laser. The ($P_1$,~$P_2$) spatial distributions given in figures~\ref{fig_pol_maps}(v3,4) and (h3,4) show that the polarization degree of the scattered $\gamma$-ray beam incident on the imaging plate is higher than 99$\%$.  

\subsection{Electron beam phase space distribution influence}
\label{sub_sec_emittance}

\begin{figure}[h!]  % \columnwidth
\centering
\includegraphics[width=0.95\columnwidth, angle=0]{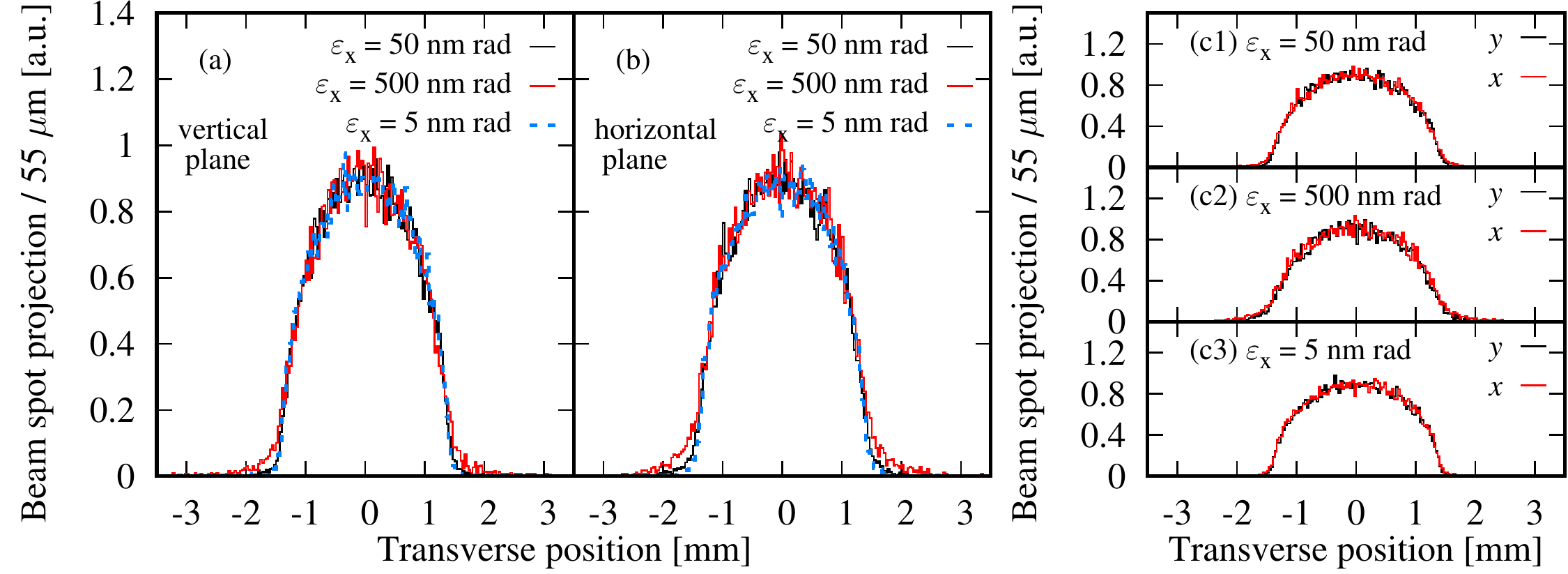}  %{ETC/FIG_EMIT/fig_emit.pdf} 
\caption{Vertical (a) and horizontal (b) projections of simulated spatial distributions for NewSUBARU LCS $\gamma$-ray beams considering the INAZUMA laser of $\lambda$~=~1.064~$\mu$m and 982~MeV electron beams of ($\varepsilon_x$, $\varepsilon_y$)~=~(50~nm$\cdot$rad, 5~nm$\cdot$rad) (black lines), ($\varepsilon_x$, $\varepsilon_y$)~=~(500~nm$\cdot$rad, 50~nm$\cdot$rad) (red lines) and ($\varepsilon_x$, $\varepsilon_y$)~=~(5~nm$\cdot$rad, 0.5~nm$\cdot$rad) (dotted blue lines) emittance values. The vertical and horizontal projections for each emittance configuration are compared in (c1~--~3).
}
\label{fig_emit}       % Give a unique label
\end{figure}

Figure~\ref{fig_emit} shows simulations of collimated LCS $\gamma$-ray beams spatial distributions for different transverse emittance parameters. We considered:
\begin{itemize}
\item $realistic$ electron beam conditions with ($\varepsilon_x$, $\varepsilon_y$)~=~(50,~5)~nm$\cdot$rad emittance,
\end{itemize}
as well as two extreme scenarios: 
\begin{itemize}
\item $bad$ electron beam conditions with ($\varepsilon_x$, $\varepsilon_y$)~=~(500,~50)~nm$\cdot$rad emittance,
\item and nearly $ideal$ electron beam conditions with ($\varepsilon_x$, $\varepsilon_y$)~=~(5,~0.5)~nm$\cdot$rad emittance.  
\end{itemize}

Figures~\ref{fig_emit}(a) and (b) show vertical and respectively horizontal projections of LCS $\gamma$-ray beams spatial distributions for the three electron beam conditions. Although there is little difference between results obtained with the $realistic$ and $ideal$ electron beams, for $bad$ electron beam conditions we notice a tail increase both in the vertical and in the horizontal projections. Figures~\ref{fig_emit}(c1--3) show that the vertical and horizontal beam spot projections are nearly identical for all electron beam conditions, with slightly higher tails for the horizontal projection. A significant incident flux decrease with the deterioration of the electron beam conditions is observed from the lower statistics of the (500,~50)~nm$\cdot$rad emittance results, as the same number of events has been used for the three simulated scenarios. 

The spatial distribution projections shown in figure~\ref{fig_emit} are scaled to their respective peak values. The three simulated scenarios have the same number of initial sampled events. The lower statistics observed for the (500,~50)~nm$\cdot$rad emittance results is due to the significant incident flux decrease with the deterioration of the electron beam conditions. The $bad$ (500,~50)~nm$\cdot$rad and the $ideal$ (5,~0.5)~nm$\cdot$rad emittance scenarios beam flux values are $\sim$56$\%$ and respectively $\sim$118$\%$ of the incident flux for the $realistic$ (50,~5)~nm$\cdot$rad emittance scenario. 

\subsection {Electron beam energy influence}
\label{sub_sec_en_ele}

\begin{figure}[h!]  % \columnwidth
\centering
\includegraphics[width=0.95\columnwidth, angle=0]{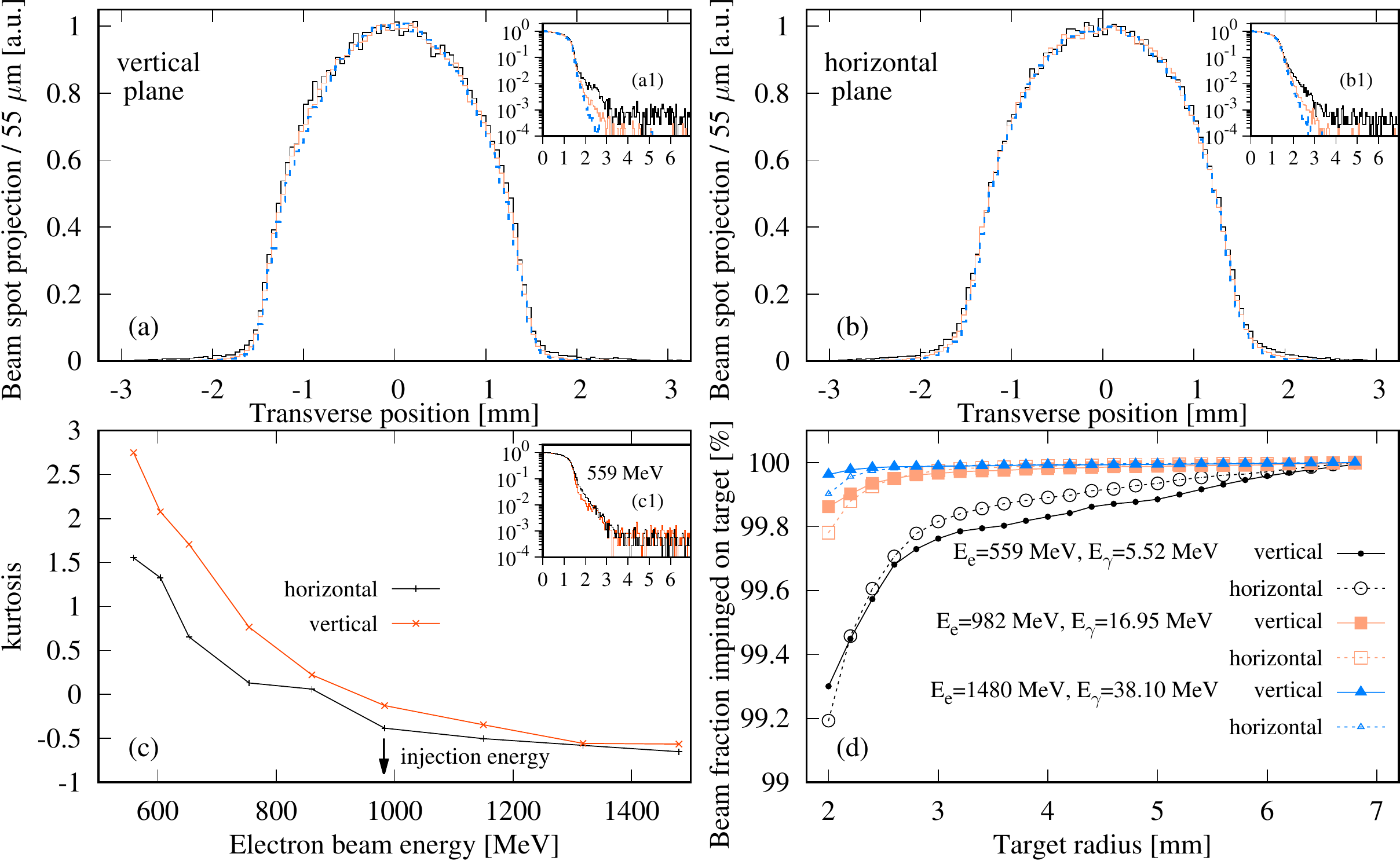}   %{ETC/FIG_EN_STUDY/fig_en_sim.pdf} % %
\caption{Simulations of collimated LCS $\gamma$-ray beam spatial distribution transverse projections and related characteristics. (a) Vertical and (b) horizontal plan projections obtained considering the 1064~nm wavelength INAZUMA laser and 559~MeV (black lines), 982~MeV (red lines) and 1480~MeV (blue lines) electron beam energies. The beam halo is investigated by plotting the (a1) vertical and (b1) horizontal projections in logarithmic scale. (c) The horizontal (black) and vertical (red) beam kurtosis versus the electron beam energy. The transverse profile projections for the 559~MeV electron energy are displayed in (c1) for understanding the difference between the vertical and horizontal kurtosis values. (d) The percentage of the LCS $\gamma$-ray beam flux impinged on a circular target versus the target radius.  
}
\label{fig_en_sim}       % Give a unique label
\end{figure}

Figures~\ref{fig_en_sim}(a) and (b) show simulations for vertical and horizontal projections of spatial distributions for collimated LCS $\gamma$-ray beams, where we considered the 1064~nm wavelength INAZUMA laser and three different electron beam energies:
\begin{itemize}
\item $E_e$~=~559~MeV, the lower limit of the electron beam energy range, corresponding to $E_\gamma$~=~5.52~MeV;
\item $E_e$~=~982~MeV, the injection energy, corresponding to $E_\gamma$~=~16.95~MeV;
\item $E_e$~=~1480~MeV, the upper limit of the electron beam energy range, corresponding to $E_\gamma$~=~38.10~MeV. 
\end{itemize}

We notice that the collimated LCS $\gamma$-ray beam core, represented by the peak region of the transverse spatial projections, is insensitive to the electron beam energy. However, the beam halo, shown in figures~\ref{fig_en_sim}(a1) and (b1) by the side tails in the spatial projections, grows with decreasing the electron beam energy. We notice in figures~\ref{fig_en_sim}(a1) and (b1) that for the low 559~MeV electron beam energy, the halo profile consists of a shoulder in the core vicinity and a rather flat contribution well outside the core radius.

Figure~\ref{fig_en_sim}(c) shows the vertical and horizontal beam excess kurtosis, a quantity defined in terms of the fourth moment of the distributions and thus sensitive to the distribution tails. The reference standard is the normal distribution, for which the kurtosis is 0. The kurtosis decreases as a profile becomes more flat, or square-like and increases as the tails increase. Figure~\ref{fig_en_sim}(c) shows that, in the low electron beam energy region between 559~and~982~MeV, the kurtosis decreases from a rather high positive value of $\sim$3 to 0. Above the electron beam injection energy, the kurtosis is negative and shows a slow decrease with increasing the electron energy. We notice that the vertical kurtosis is slightly higher than the horizontal one. This is explained in figure~\ref{fig_en_sim}(c1) by plotting in logarithmic scale the vertical (red) and horizontal (black) projections of the spatial distribution for the lowest investigated electron beam energy. We notice that, in the core vicinity, the vertical projection tails fall more rapidly than the horizontal ones but, well outside the core are higher than the horizontal ones.

The fraction of incident photons present the surrounding halo is however small. Figure~\ref{fig_en_sim}(c) shows the percentage of the collimated LCS $\gamma$-ray beam flux impinged on a circular target with radius in the 2~mm to 7~mm range, as obtained by integrating the vertical and horizontal spatial projections. Here we investigated the lower (559~MeV) and upper (1480~MeV) electron beam energy limits as well as the injection energy of 982~MeV. The simulations show that more than 99$\%$ of the incident flux is impinged on a 2~mm radius target. The radii of targets used in the NewSUBARU photoneutron measurements range between 4~mm and 10~mm, assuring that more than 99.8$\%$ of the incident flux hits the target surface. 

\section{Probing spatial distributions with a MiniPIX X-ray camera}
\label{sec_exp_distributions}

Experimental collimated bremsstrahlung and LCS $\gamma$-ray beam spatial distributions have been recorded with a MiniPIX X-ray camera~\cite{minipix_website,granja_2022}. The MiniPIX device consists of a 500~$\mu$m thick Silicon wafer with 256~$\times$~256 pixels. The pixel pitch is 55~$\mu$m. The MiniPIX camera has been placed in the same position as the virtual imaging plate considered in section~\ref{sec_sp_distrib}, at 25.5~m downstream of the BL01 beamline midpoint, corresponding to the target position in photoneutron experiments. A 0.2~mm thick stainless steel protective cover was placed in front of the MiniPIX camera during the measurements. The experimental spatial distributions are compared with \texttt{eliLaBr} simulation results. 

\subsection{MiniPIX 2D histograms}
\label{sub_sec_minipix}

\begin{figure*}[h!]  % \columnwidth
\centering
\includegraphics[width=0.95\textwidth, angle=0]{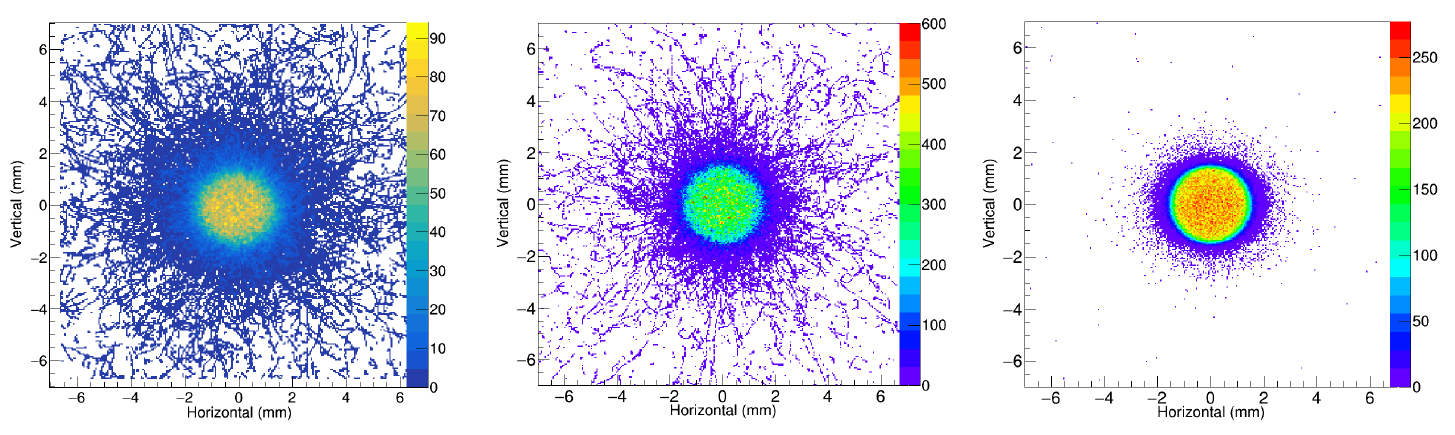} %{ETC/FIG_beamspot2D/fig_beamspot_2D.pdf} 
\put(-400,125){\small{(a) MiniPIX - experimental}}
\put(-280,125){\small{(b) Energy deposition - simulation}}
\put(-140,125){\small{(c) Incident beamspot - simulation}}
\caption{Spatial distribution of collimated LCS $\gamma$-ray beam produced with the INAZUMA laser and 982~MeV electron beam: (a) experimental MiniPIX histogram, (b) simulated energy deposition in the MiniPIX detector and (c) simulated spatial distribution incident on the MiniPIX detector.   
}
\label{fig_beamspot_2D}       % Give a unique label
\end{figure*}

Figure~\ref{fig_beamspot_2D}(a) shows the 2D spatial profile recorded by the MiniPIX camera for a $E_\gamma$~=~16.95~MeV maximum energy LCS $\gamma$-ray beam produced using the INAZUMA laser and $E_e$~=~982~MeV electron beams. Figures~\ref{fig_beamspot_2D}(b) and (c) show the corresponding \texttt{eliLaBr} simulations. In order the reproduce the experimental measurements, the histogram shown in figure~\ref{fig_beamspot_2D}(b) is constructed by incrementing the energy deposition events in the 0.5~mm thick Silicon plate. Figure~\ref{fig_beamspot_2D}(c) shows the spatial distribution of the photon beam impinged on the imaging plate. 

We notice that both the experimental and the simulated energy deposition histograms show a much wider halo than the incident distribution. The wide halo it generated by the electromagnetic interaction of the incident beam inside the MiniPIX detector, which can be observed in figures~\ref{fig_beamspot_2D}(a) and (b) as the visible particle traces.

\subsection{Bremsstrahlung background}
\label{sub_sec_bremsstrahlung}

We further investigate the contribution of the bremsstrahlung background produced by the relativistic electron beams circulating in the storage ring. Although its energy spectrum decreases rapidly with the photon energy (see figure~6 of ref.~\cite{aoki2004} for an experimental NewSUBARU bremsstrahlung spectrum), bremsstrahlung is a main source of beam-related background. 

In time projection chamber experiments, are discarded during offline particle track analysis~\cite{gros_2018,gai_2010}. In slow-response, moderated neutron detection array experiments, the amount of bremsstrahlung-induced neutrons is subtracted by performing separate bremsstrahlung~+~beam and only-bremsstrahlung measurement~\cite{Gheorghe2017,silano_2018}. Here we note the $^{238}$U photofission cross section measurements of Csige~{\it et al.}~\cite{csige_2013} and of J.A.~Silano and H.J.~Karwowski~\cite{silano_2018}. Here the low-energy discrepancies between the two data sets have been attributed to an underestimation of the bremsstrahlung background component in the former experiment. 

Thus, we investigated the bremsstrahlung effect on the MiniPIX spatial distribution measurements. The experimental investigation is compared with \texttt{eliLaBr} simulations,  which now include bremsstrahlung model. The present code simulates the interaction of electrons with air molecules inside the 20~m straight section of the electron beamline. 

\begin{figure*}[h]  % \columnwidth
\centering
\includegraphics[width=0.95\textwidth, angle=0]{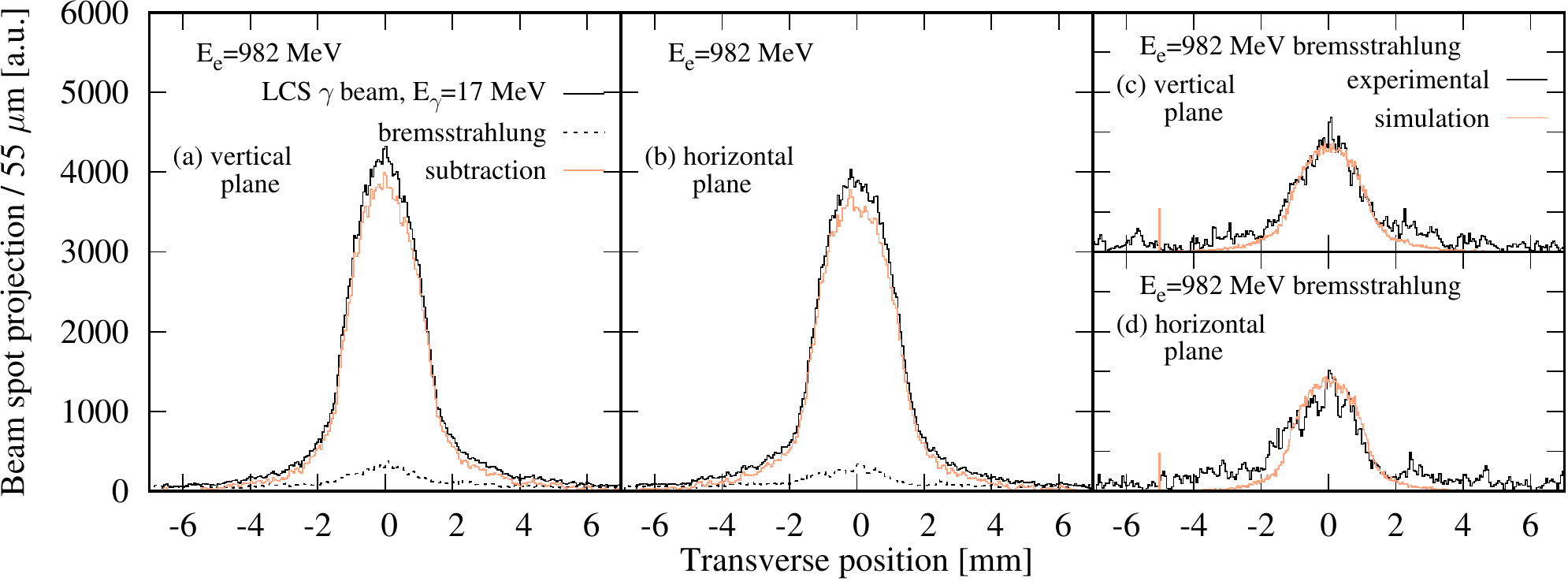} 
\caption{Collimated bremsstrahlung and LCS $\gamma$-ray beam spatial distributions for $E_e$~=~982~MeV and the INAZUMA laser. The (a) vertical and (b) horizontal projections of the MiniPIX experimental beamspot results are shown for the LCS $\gamma$ beam plus bremsstrahlung (solid black lines), for bremsstrahlung-only (dotted black line) and for the background subtracted LCS $\gamma$-ray beam (orange lines). The bremsstrahlung (c) vertical and (d) horizontal MiniPIX beamspot projections (black lines) are compared with \texttt{eliLaBr} simulations (orange lines).   
}
\label{fig_brems}       % Give a unique label
\end{figure*}

Experimental (a) vertical and (b) horizontal projections of collimated bremsstrahlung-only (dotted black lines) and LCS $\gamma$-ray beam plus bremsstrahlung (solid black lines) are shown in figure~\ref{fig_brems}. For the measurement, 300~mA electron beams at the $E_e$~=~982~MeV injection energy and a double collimation system of C1~=~3~mm and C2~=~2~mm have been used. The MiniPIX irradiations lasted for 600 seconds each. We notice that the bremsstrahlung component is a small fraction of LCS $\gamma$-beam, accompanied with bremsstrahlung. The LCS $\gamma$-ray beam-only component, obtained as the direct subtraction of the two measurements, is also displayed in orange lines. The experimental bremsstrahlung beam spot projections are well described by the \texttt{eliLaBr} simulations, as shown in figures~\ref{fig_brems}(c,~d).   

\begin{figure*}[h]  % \columnwidth
\centering
\includegraphics[width=0.95\textwidth, angle=0]{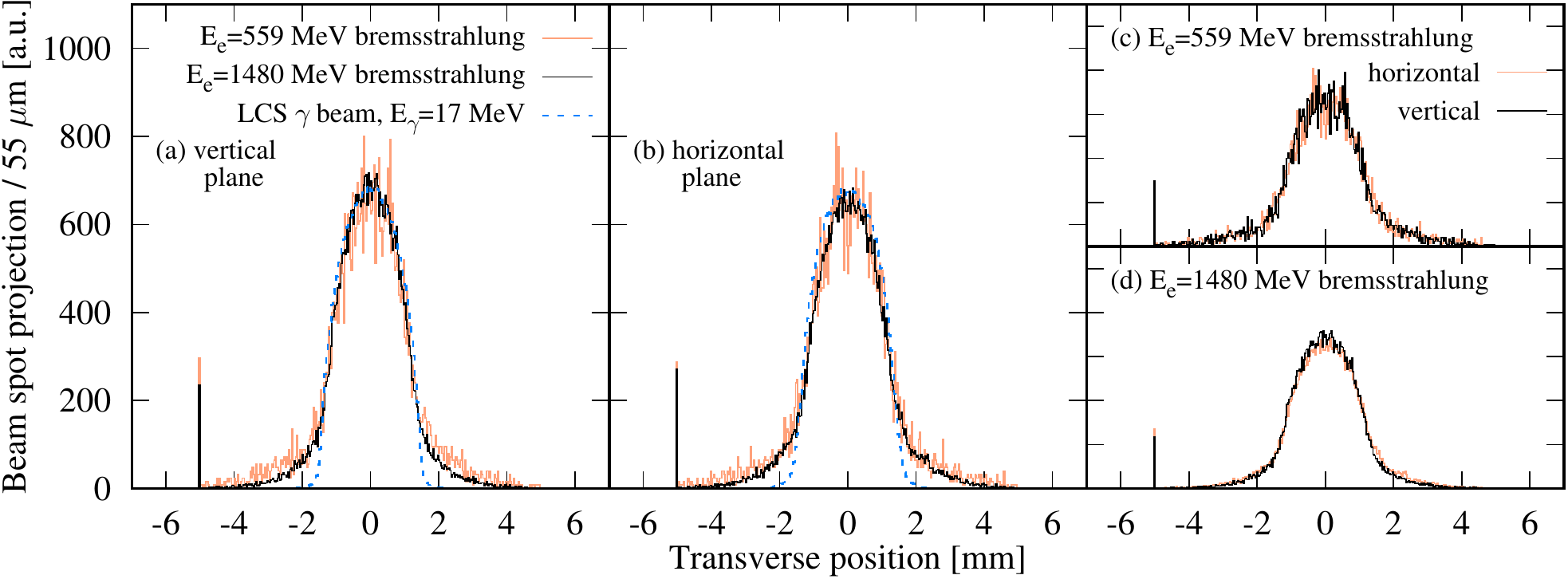} 
\caption{Collimated bremsstrahlung and LCS $\gamma$-ray beams spatial distributions for different electron beam energies, as obtained by the \texttt{eliLaBr} simulations. The (a) vertical and (b) horizontal beamspot projections for bremsstrahlung-only beams produced with $E_e$~=~559~MeV (orange lines) and with $E_e$~=~1480~MeV (solid black lines) are compared with $E_\gamma$~=~16.95~MeV LCS $\gamma$-ray beam ones (dashed blue lines). The horizontal (orange lines) and vertical (black lines) beamspot projections are compared  for the (c) $E_e$~=~559~MeV and (d) $E_e$~=~1480~MeV bremsstrahlung beams.  
}
\label{fig_brems_var_ene}       % Give a unique label
\end{figure*}

We further investigate the spatial distribution of collimated bremsstrahlung beams produced with different electron beam energies. Figure~\ref{fig_brems_var_ene} shows the simulated (a) vertical and (b) horizontal beamspot projections for the  NewSUBARU lower and upper limit electron beam energy range, of 559~MeV (orange lines) and respectively 1480~MeV (black lines). 

First, from the significantly poorer statistics of the 559~MeV projections compared with the 1480~MeV ones, we notice a strong flux decrease with the energy decrease, as both simulations were performed for the same number of initial uncollimated sampled events. Here, the 559~MeV projections are multiplied by a factor of 8 in order to compare them to the 1480~MeV ones.

Second, we notice that the halo slightly increases with decreasing the electron beam energy. However, both bremsstrahlung beamspot projections present a significantly larger halo than the collimated LCS $\gamma$-ray beam halo shown by dashed blue lines in figures~\ref{fig_brems_var_ene}(a,~b).  

The horizontal and vertical beamspot projections are compared in figures~\ref{fig_brems_var_ene}(c) for $E_e$~=~559~MeV and in (d) for $E_e$~=~1480~MeV. For the more productive $E_e$~=~1480~MeV bremsstrahlung beam, we can observe that the vertical projection is slightly narrower than the horizontal one, with a higher maximum and lower halo tails. 

\begin{figure*}[h]  % \columnwidth
\centering
\includegraphics[width=0.85\textwidth, angle=0]{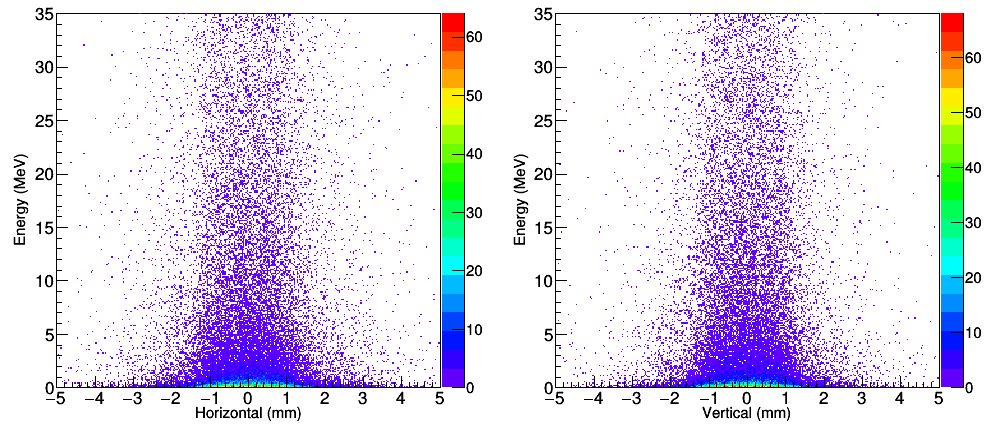} 
\put(-335,130){(a)}
\put(-155,130){(b)}
\caption{Collimated bremsstrahlung energy spatial distributions on (a) horizontal and (b) vertical axes at the target position. Simulations have been performed for $E_e$~=~1480~MeV electron beams.   
}
\label{fig_Ene_1460_ene_sp_distrib_crop}       % Give a unique label
\end{figure*}

Figure~\ref{fig_Ene_1460_ene_sp_distrib_crop} shows simulations of collimated bremsstrahlung energy spatial distributions in the (a) horizontal and (b) vertical planes for $E_e$~=~1480~MeV electron beams. We notice that the high energy component of the spectrum is present in the beam core, while the beam halo contains mostly low energy photons below 1~MeV. 

\subsection{LCS $\gamma$-ray beams with the INAZUMA and the CO$_2$ laser}
\label{sub_sec_minipix_LCS}

We further validate the \texttt{eliLaBr} simulations results given in section~\ref{sec_sp_distrib} against NewSUBARU experimental spatial distribution data. A double collimation system of C1~=~3~mm and C2~=~2~mm has been used, the same as in the LCS simulations shown in section~\ref{sec_sp_distrib} and the bremsstrahlung ones of section~\ref{sub_sec_bremsstrahlung}. 

LCS $\gamma$-ray beams were produced in scatterings of 1064~nm INAZUMA laser photons on relativistic electrons at energies between 559~MeV and 1480~MeV and of 10591~nm CO$_2$ laser photons on 982~MeV and 1480~MeV electron beams. Typical operating conditions for the NewSUBARU LCS $\gamma$-ray beam line were used: relativistic electron currents between 70 and 300 mA and laser power between of 1 and 6 W, which generated a rate of $\sim$10$^5$ $\gamma$ photons per second. Spatial profiles of all LCS $\gamma$-ray beams were recorded with the MiniPIX X-ray camera, where a acquisition time of 10~minutes was set for each measurement. Thus, we tested the MiniPIX imaging camera for beamspot monitor applications in a wide LCS $\gamma$-ray energy range, between 1.73~MeV and 38.1~MeV. 

\begin{figure*}[h!]  % \columnwidth
\centering
\includegraphics[width=0.95\textwidth, angle=0]{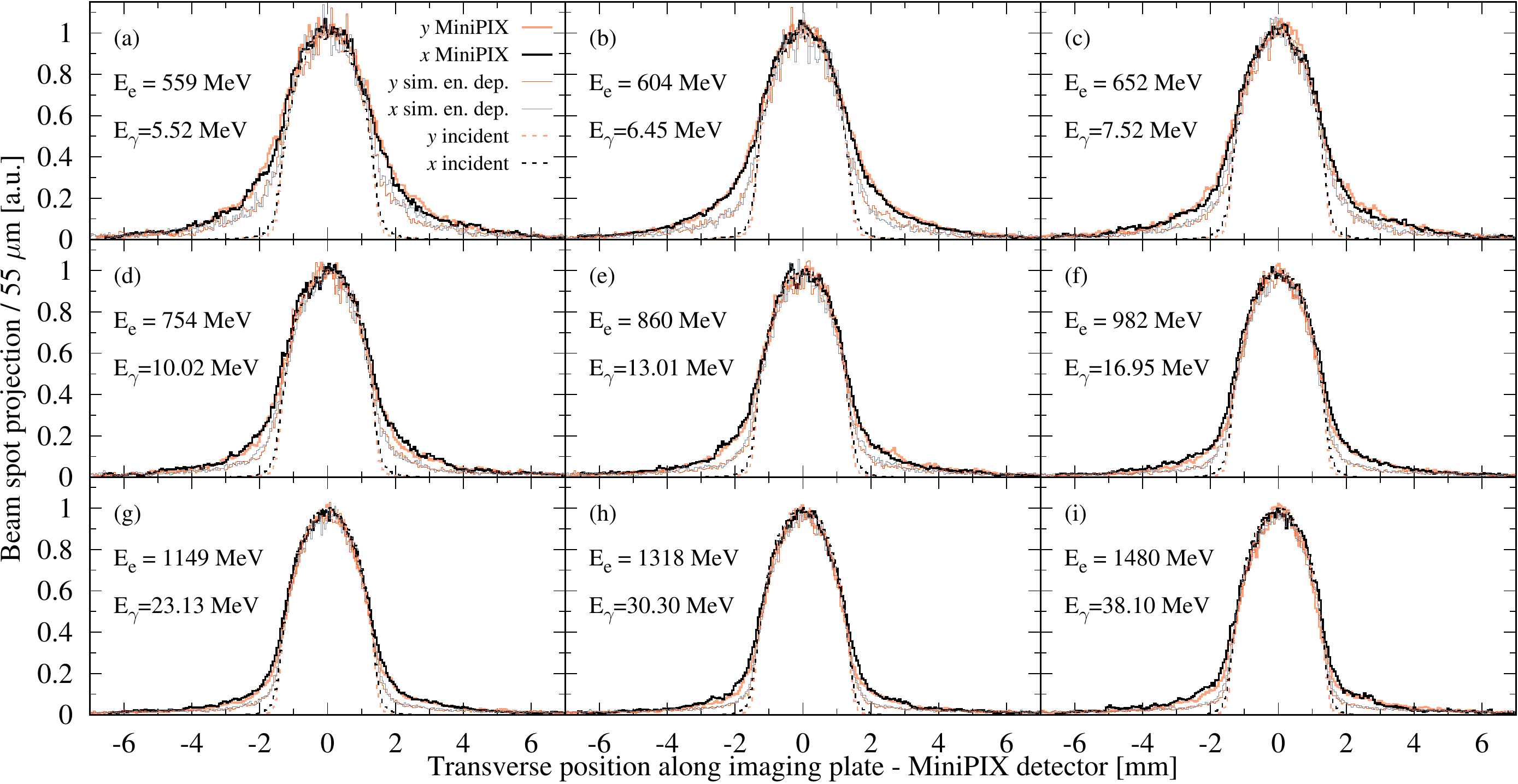} %{ETC/FIG_EN_STUDY/fig_en_exp.pdf} 
\caption{Spatial distributions for collimated LCS $\gamma$-ray beams produced with INAZUMA laser and electron beams of (a) 559~MeV, (b) 604~MeV, (c) 652~MeV, (d) 754~MeV, (e) 860~MeV, (f) 982~MeV, (g) 1149~MeV, (h) 1318~MeV and (i) 1480~MeV. Vertical (red) and horizontal (black) projections for the experimental MiniPIX detector response (thick solid lines), for the simulated energy deposition distributions (thin solid lines) and for the simulated incident distributions (dotted lines).    
}
\label{fig_en_exp}       % Give a unique label
\end{figure*}

\begin{figure*}[h!]  % \columnwidth
\centering
\includegraphics[width=0.95\textwidth, angle=0]{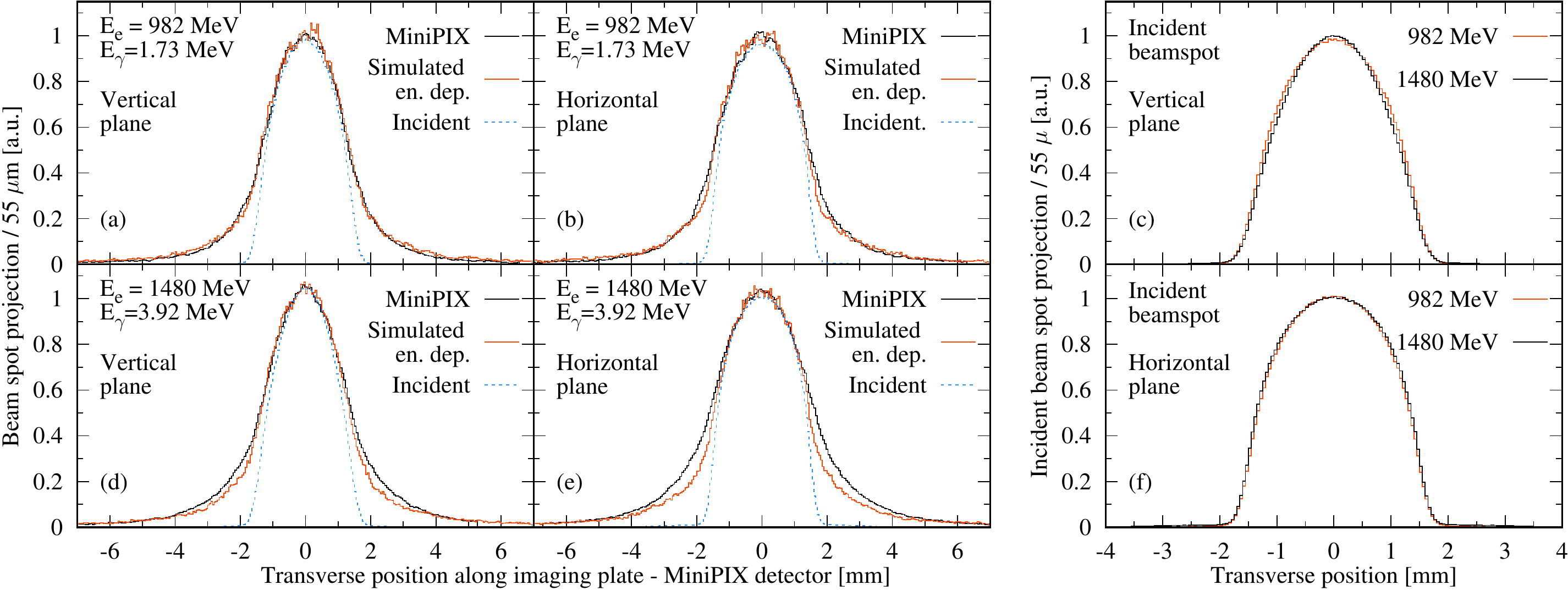} %{ETC/FIG_CO2/fig_co2_exp_sim.pdf} 
\caption{Spatial distributions for collimated LCS $\gamma$-ray beams produced with CO$_2$ laser and electron beams of 982~MeV and 1480~MeV, as following. The experimental MiniPIX (black lines) and the simulated energy deposition (red) and incident (dotted blue) spatial distributions are compared in (a) vertical and (b) horizontal projections for the 982~MeV, (d) vertical and (e) horizontal projections for 1480~MeV. The (c) vertical and (f) horizontal projections of the incident spatial distributions are also compared for the two electron beam energies.       
}
\label{fig_co2_exp_sim}       % Give a unique label
\end{figure*}

Figure~\ref{fig_en_exp} and \ref{fig_co2_exp_sim} show the experimental and simulated spatial distribution projections in the vertical and horizontal plane for the LCS $\gamma$-ray beams obtained with the INAZUMA and the CO$_2$ laser, respectively. The experimental beamspot projections are well reproduced by the \texttt{eliLaBr} simulations of the spatial distribution of energy deposition in the Silicon plate.

As observed also in section~\ref{sub_sec_minipix}, the experimental MiniPIX projections and the simulated energy deposition ones show a significantly wider halo than the incident LCS $\gamma$-ray beam projections. The difference originates from a shower of secondary particles produced by the electromagnetic interaction of the incident LCS $\gamma$-rays with the 0.2~mm thick stainless protective cover, and the electromagnetic interaction of the incident photon beam in the MiniPIX plate. The stainless steel protective cover has been used to prevent the electrons and photons resulted from beam scatterings on air molecules and collimators to enter the MiniPIX detector. However, we found that, for low energy LCS $\gamma$-ray beams, the secondary radiation generated in the cover brings significant contribution to the total beam spot measurement and therefore broadens the total beam spot comparing to the pure LCS $\gamma$-beam spot. Thus, we conclude that the MiniPIX should be used without the protective cover for LCS $\gamma$-ray spatial distribution measurements. We also notice that for all incident energies, the tails of the experimental MiniPIX distributions are slightly higher than the simulated ones. This is given by the fact that no energy resolution, charge collection effects or cross talk between neighboring pixels have been considered in the \textsc{Geant4} simulations.  

\section{Conclusions}
\label{sec_conclusions}

The high intensity, small beamspot size, small divergence and small bandwidth qualify collimated laser Compton scattering $\gamma$-ray beams as suitable tools for nuclear applications with costly target materials. Thus, the accurate description and prediction of the LCS $\gamma$-ray beam spatial distribution is a key aspect for applications such as the production of medical radioisotopes and absolute measurements of photonuclear reactions cross sections. 

It is well known that the collimated LCS $\gamma$-ray beam core is defined by the collimator system. In this paper, we report our investigation of LCS $\gamma$-ray beam spatial distribution as function of the electron beam energy, electron beam phase space distribution, laser optics conditions and laser polarization. We have found that variations in the above listed parameters affect the LCS $\gamma$-ray beam halo to different extents.

We show that the electron beam phase space deterioration and the decrease in the electron beam energy generate a growth in the $\gamma$-beam halo. We have shown that although the polarization state of the laser influences on the $\gamma$-ray beam energy distribution, it does not affect the $\gamma$-ray beam spatial distribution. We have also found that the spatial distribution of the bremsstrahlung beam is wider than the LCS $\gamma$-ray beam one and that it also increases with the electron beam energy decreasing.

For the present study, we used LCS $\gamma$-ray beam simulations produced with the \texttt{eliLaBr} code and real collimated LCS $\gamma$-ray beams produced at the NewSUBARU synchrotron radiation facility. We have successfully tested a 500~$\mu$m MiniPIX X-ray camera as a beamspot monitor in a wide $\gamma$-ray beam energy range between 1.73~MeV and 38.1~MeV. 

\acknowledgments

D.F. acknowledges the support from the Romanian Ministry of Research, Innovation and Digitalization/Institute of Atomic Physics from the National Research - Development and Innovation Plan III for 2015-2020/Programme 5/Subprogramme 5.1 ELI-RO, project GANT-Photofiss No 14/16.10.2020. 
This work was supported by a grant of the Ministry of Research, Innovation and Digitization, CNCS - UEFISCDI, project number PN-III-P1-1.1-PD-2021-0468, within PNCDI III.

% We suggest to always provide author, title and journal data:
% in short all the informations that clearly identify a document.

\end{document}